\documentclass[%
reprint,
superscriptaddress,
 amsmath,amssymb,
 aps,
floatfix,
]{revtex4-1}

\usepackage{graphicx}
\usepackage{dcolumn}
\usepackage{bm}
\usepackage[dvipdfmx]{hyperref}
\hypersetup{
	colorlinks=true,
	linkcolor=blue,          
	citecolor=blue,        
	filecolor=blue,
	urlcolor=blue
}

\begin{document}

\preprint{APS/123-QED}

\title{Enhanced violation of the Collins-Gisin-Linden-Massar-Popescu inequality \\ with optimized time-bin-entangled ququarts}

\author{Takuya Ikuta}
 \email{ikuta@procyon.comm.eng.osaka-u.ac.jp}
 \affiliation{%
 NTT Basic Research Laboratories, NTT Corporation, 3-1 Morinosato Wakamiya, Atsugi, Kanagawa 243-0198, Japan}
\affiliation{Division of Electrical, Electronic and Information Engineering, Osaka University, Suita, Osaka 565-0871, Japan
}%
\author{Hiroki Takesue}
\affiliation{%
 NTT Basic Research Laboratories, NTT Corporation, 3-1 Morinosato Wakamiya, Atsugi, Kanagawa 243-0198, Japan}

\date{\today}

\begin{abstract}
High-dimensional quantum entanglement is drawing attention
because it enables us to perform quantum information tasks that are robust against noises.
To test the nonlocality of entangled qudits,
the Collins-Gisin-Linden-Massar-Popescu (CGLMP) inequality has been proposed
and demonstrated using qudits based on orbital angular momentum, time-energy uncertainty, and frequency bins.
Here, we report the generation and observation of time-bin entangled ququarts.
We implemented a measurement for the CGLMP inequality test using cascaded delay Mach-Zehnder interferometers fabricated by using planar lightwave circuit technology, with which we achieved a precise and stable measurement for time-bin-entangled ququarts.
In addition, we generated an optimized entangled state by modulating the pump pulse intensities,
with which we can observe the theoretical maximum violation for the CGLMP inequality test.
As a result, we successfully observed a Bell-type parameter $S_4 = 2.774 \pm 0.025$ violating the CGLMP inequality for the maximally entangled state
and an enhanced Bell-type parameter $S_4 = 2.913 \pm 0.023$ for the optimized entangled state.
\end{abstract}

\pacs{Valid PACS appear here}
\maketitle


\section{\label{sec:Intro}Introduction}

Quantum entanglement has a strange non-local characteristic unique to quantum particles, known as the Einstein-Podolsky-Rosen paradox \cite{Einstein1935}.
Ever since the first reports of the violation of Bell's inequality \cite{Freedman1976,Aspect1981},
the non-locality of quantum entanglement has been widely confirmed,
and subsequent efforts to overcome the loopholes have been reported \cite{Ansmann2009,Giustina2013,Christensen2013,Hensen2015,Giustina2015,Shalm2015}.
Now quantum entanglement has been utilized for many quantum communication tasks such as quantum key distribution (QKD) \cite{Ekert1991},
quantum teleportation \cite{Bennett1993}, and quantum repeaters \cite{Sangouard2011,Azuma2015}.
Recently, entangled qudits, or high-dimensional entanglement, have been attracting attention,
because they enable us to perform robust quantum information tasks.
For example,
it has been theoretically shown \cite{Cerf2002,Sheridan2010} and experimentally demonstrated \cite{Groblacher2006a,Mafu2013,Zhong2015} that we can improve the error rate tolerance for QKD.
Furthermore, the amount of information per photon can be increased \cite{Zhong2015}.
In addition, we can use the entangled qudits to relax the condition for closing the detection loophole in Bell's inequality tests \cite{Vertesi2010}.

To test the nonlocality of the entangled qudits,
Collins, Gisin, Linden, Massar, and Popescu (CGLMP) proposed a Bell-type inequality\cite{Collins2002}.
A unique characteristic of the CGLMP inequality is that
non-maximally entangled states give the theoretical maximum violation of the inequality for $d>2$, where $d$ is a Hilbert space dimension of a particle \cite{Dada2011a}.
So far,
CGLMP inequality tests have been performed by using qudits based on orbital angular momentum (OAM) \cite{Vaziri2002,Dada2011a},
time-energy uncertainty \cite{Richart2012,Thew2004}, and frequency-bins \cite{Bessire2014,Bernhard2013}.
Among these, OAM-based entangled qudits have been intensively studied,
and even a $100 \times 100$ dimensional entanglement has been demonstrated \cite{Krenn2014}.
Despite their advantage for scalability,
OAM-based qudits are not suitable for the fiber transmission required for long-distance terrestrial QKD because of the spatial mode dispersion in fiber. 
Time-energy entangled photon pairs are promising candidates for fiber transmission.
However,
with time-energy entanglement,
it is difficult to generate the optimized entangled state for obtaining the maximum violation of the CGLMP inequality.
Bernhard {\textit{et al.}} \cite{Bernhard2013} attempted to generate optimized frequency-bin-entangled qudits with spatial light modulators (SLMs).
They successfully observed violations of the CGLMP inequality;
however, no significant difference between the maximally entangled qudits and the optimized entangled qudits was observed.
In addition, the optimization using SLMs causes excess loss,
which results in a lower coincidence rate in the CGLMP measurement.

Here, we report the generation and observation of time-bin-entangled ququarts, or four-dimensional entangled photon pairs.
We used cascaded delay Mach-Zehnder interferometers (MZIs) \cite{Richart2012} to implement a measurement for the CGLMP inequality test for the time-bin-entangled ququarts.
An advantage of cascaded delay MZIs is that they require fewer optical passes than multi-arm interferometers \cite{Thew2004}.
A drawback of this implementation is that it introduces more loss than the multi-arm interferometers.
Using cascaded delay MZIs made with planar lightwave circuit technology (PLC),
we observed clear two-photon interferences and a direct violation of the CGLMP inequality.
Furthermore, we generated an optimized entangled state that enabled the maximum violation of the CGLMP inequality.
We were able to modulate probability amplitudes of the time-bin-entangled ququarts to obtain the optimized entangled state
simply by modulating pump pulse intensities.
Consequently, we observed an enhanced violation of the CGLMP inequality by 39 standard deviations with the optimized entangled state.

\section{\label{sec:CGLMP} The CGLMP inequality and optimized entangled state}

The CGLMP inequality \cite{Collins2002} is a natural extension of the Clauser-Horne-Shimony-Holt (CHSH) inequality \cite{JohnF.ClauserMichaelA.HorneAbnerShimony1969}.
In the CHSH inequality,
we quantify the correlation between two photons by $\pm1$ values according to the results measured on the each side of the receivers,
Alice and Bob. With a high-dimensional system, however, there are more orthogonal states than 2,
and we thus need more values corresponding to these states.
For this purpose, the CGLMP inequality has a Bell-type parameter, which is given by
\begin{widetext}
	\begin{eqnarray}
		S_{d} = \sum_{k=0}^{[d/2]-1} \left( 1-\frac{2k}{d-1} \right)
			\left \{
				\left[
					P(A_0 = B_0 + k)
					+P(B_0 = A_1 + k+1)
					+P(A_1 = B_1 + k)
					+P(B_1 = A_0 + k)
				\right]
			\right.		\notag	\\
			\left.
				- \left[
					P(A_0 = B_0 - k -1)
					+P(B_0 = A_1 - k)
					+P(A_1 = B_1 - k -1)
					+P(B_1 = A_0 - k -1)
				\right]
			\right \}	,		\label{eq:SValueDefinition}
	\end{eqnarray}
\end{widetext}
where $d$ is the dimension and $A_a$ and $B_b$ are Alice and Bob's measurement outcomes when they select the bases $a$, $b \in \{0,1\}$, respectively.
$P(X_x = Y_y + Z)$ is a conditional probability that $X_x$ is equal to $(Y_y + Z)$ modulo $d$
when measurement bases are set at $x$ and $y$,
where $X_x$, $Y_y \in \{ A_a, B_b \}$, $x$, $y \in \{ a, b \}$, and $Z \in \mathbb{Z}$.
In the local hidden variable theory, $S_d$ satisfies the following inequality as with the CHSH inequality \cite{Collins2002}:
\begin{equation}
	S_d \leq 2, \textnormal{for any} \ d \geq 2	.		\label{eq:CGLMPinequality}
\end{equation}
However, this inequality is violated using quantum entanglement;
therefore, the violation of the CGLMP inequality is good evidence of a high-dimensional entanglement.

For this test, one can use the maximally entangled state, which is given by
\begin{equation}
	\left | \Psi _ {\textnormal{MES}} \right > = \frac{1}{\sqrt{d}} \sum_{k=0}^{d-1} \left | k \right >_A \otimes \left | k \right >_B	,	\label{eq:MaxEnt}
\end{equation}
where $\left | k \right >_A$ and $\left | k \right >_B$ are states that photons for Alice and Bob exist in the $k$-th time slot, respectively.
Using the maximally entangled state, we can observe the violation of the CGLMP inequality.
In the two-dimensional system,
the violation obtained with the maximally entangled state is the theoretical maximum violation for any state.
In a high-dimensional system, however,
non-maximally entangled states can give the maximum violation. 
In \cite{Dada2011a}, Dada \textit{et al.} derived the optimized entangled state to obtain the maximum violation.
The optimized entangled state is an eigenstate of an operator representing the measurement of $S_d$.
For example, the optimized entangled ququarts are given by
\begin{equation}
	\left | \Psi _ {\textnormal{OES}} \right> = \frac{1}{\sqrt{2 (1+ \gamma^2 ) }} \left( \left| 0,0 \right> + \gamma \left| 1,1 \right> + \gamma \left| 2,2 \right> + \left| 3,3 \right>  \right)	,	\label{eq:OptEnt}
\end{equation}
where $\gamma \approx 0.739$ is an amplitude modulation factor and $\left| k,k \right>$ is a short form of $ \left | k \right >_A \otimes \left | k \right >_B$ for simplicity.
Table \ref{Tab:Vio_Max_Opt} shows the theoretical maximum $S_d$ values
for the maximally entangled states and the optimized entangled states for various $d$ values derived as in \cite{Dada2011a}.
The theoretical maximum $S_4$ for the optimized entangled state, $S_4 = 2.9727$,
is larger than the theoretical maximum $S_4$ for the maximally entangled state, $S_4 = 2.8962$.

\begin{table}[htb]
	\caption{Theoretical maximum $S_d$ for $d$-dimensional maximally entangled states and optimized entangled states.}
	\begin{tabular}{ccc}	\hline \hline
		$d$	&	$\max S_d$ for $\left | \Psi _ {\textnormal {MES}} \right >$	&	$\max S_d$ for $\left | \Psi _ {\textnormal{OES}} \right>$	\\ \hline
		2 & 2.8284 & 2.8284	\\
		3 & 2.8729 & 2.9149	\\
		4 & 2.8962 & 2.9727	\\
		\vdots & \vdots & \vdots	\\	\hline
	\end{tabular}
	\label{Tab:Vio_Max_Opt}
\end{table}

\begin{figure}[htbp]
	\begin{center}
		\includegraphics[clip,width=8.5cm]{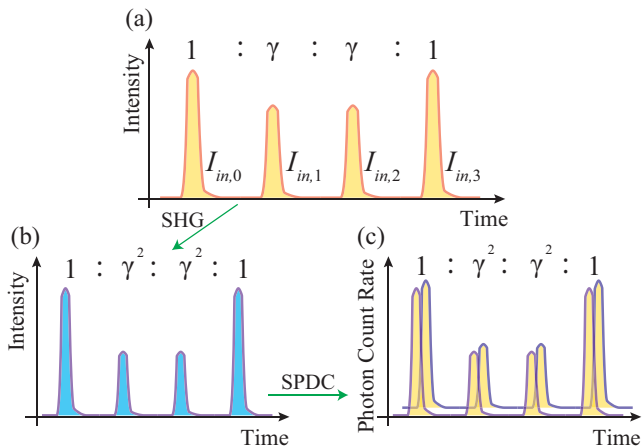}
		\caption{Concept of the optimized time-bin-entangled ququarts generation.
		(a) Intensities of the input light.
		(b) Intensities of the pump pulses generated through SHG.
		(c) Photon count histogram of the photon pairs generated via SPDC.}
		\label{fig:PulseModulation}
	\end{center}
\end{figure}
In our experiment, we generate the time-bin-entangled state by utilizing spontaneous parametric down conversion (SPDC)
pumped by pulses generated via second harmonic generation (SHG).
Through the SHG process, the field amplitude of generated light $E_{SHG}$
is proportional to the intensity of input light $I_{in}$.
The probability amplitude of a photon pair state generated via SPDC $C_{SPDC}$ satisfies $C_{SPDC} \propto E_{SHG}$.
We used these relations to generate the optimized time-bin-entangled ququarts.
Figure \ref{fig:PulseModulation} shows the concept of the optimized time-bin-entangled ququarts generation.
We modulate a classical lightwave into four-sequential pulses so that the intensities of the four-sequential pulses satisfy the following equation:
$ I_{in,0} : I_{in,1} : I_{in,2} : I_{in,3} = 1 : \gamma : \gamma : 1$,
where $I_{in,k}$ is the intensity of light in the $k$-th time slot.
Then, the generated time-bin-entangled state $\left| \Psi \right> = \sum_{k=0}^{3} c_{k} \left| k,k \right>$ satisfies
\begin{equation}
	c_0 : c_1 : c_2 : c_3 = 1 : \gamma : \gamma : 1	.		\label{eq:OptCondition}
\end{equation}
Thus, with the time-bin-entangled qudits,
we can obtain the optimized entangled state simply by modulating the amplitudes of the pump pulses without excess loss.

\section{\label{sec:MZI}Measurement for the CGLMP inequality}

\begin{figure*}[htbp]
	\begin{center}
		\includegraphics[clip,width=15.5cm]{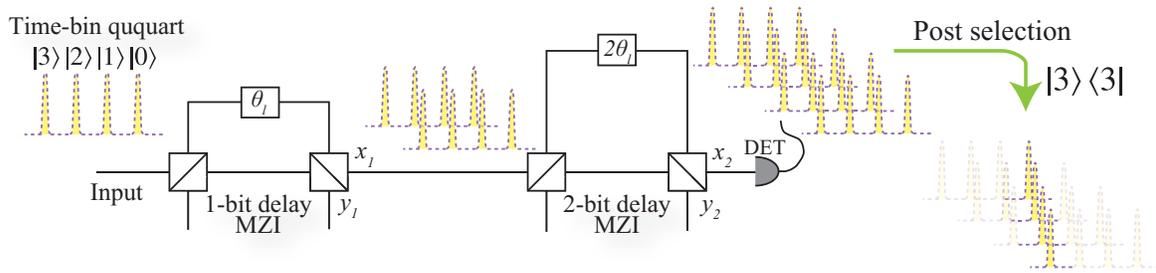}
		\caption{Concept of measurement using cascaded delay MZIs for four dimensions.}
		\label{fig:MZI}
	\end{center}
\end{figure*}
In the CGLMP inequality test, we perform a measurement projecting onto the Fourier transform basis.
The Fourier transform basis is composed of sates $\left| \theta_l \right>$, which are given by
\begin{equation}
	\left| \theta_l \right> = \frac{1}{\sqrt{d}} \sum_{k=0}^{d-1} \exp \left(i \theta_l k \right) \left| k \right>		.	\label{eq:Fourier}
\end{equation}
Phase $\theta_l$ in Eq. (\ref{eq:Fourier}) takes
\begin{equation}
	\theta_l = \frac{2 \pi}{d} \times
		\begin{cases}
			\left( l + \alpha_a \right) & {\textnormal{for Alice's basis}}\ a	\\
			\left( -l + \beta_b \right) & {\textnormal{for Bob's basis}}\ b
		\end{cases}		,	\label{eq:PhaseSet}
\end{equation}
where $\alpha_0 = 0$, $\alpha_1 = 1/2$, $\beta_0 = 1/4$, $\beta_1 = -1/4$,
and $l \in [0,d-1]$ is an integer that denotes the measurement outcomes.
We use the cascaded delay MZIs to perform this measurement.
Note that a similar scheme based on cascaded MZIs was proposed in \cite{Richart2012}.
However,
a measurement corresponding to Eq. \eqref{eq:Fourier} has not been implemented using the scheme yet.
Figure \ref{fig:MZI} shows the concept of the measurement using the cascaded MZIs for four dimensions.
The first MZI has a time delay $T$ and a phase shift $\theta_l$,
where $T$ denotes the temporal interval between time bins and $\theta_l$ the phase difference between the short and long arms.
On the other hand, the second MZI has a time delay $2T$ and a phase shift $2\theta_l$.

Here, we define a generalized measurement operator ${\cal M}_1$,
which expresses the function of the first MZI at port $x_1$ in Fig. \ref{fig:MZI} as
\begin{equation}
	{\cal M}_1 = \sum_{k=0}^3 \frac{1}{2} \left( \left| k \right> + e^{i\theta_l} \left| k + 1 \right> \right)\  \left< k \right|		.
\end{equation}
A complementary operator ${\cal M}_{1}'$ corresponding to the function of the first MZI at port $y_1$
is defined so as to satisfy ${\cal M}_1^\dag {\cal M}_1 + {\cal M}_1'^\dag {\cal M}_1' = \mathbb{I}$,
where $\mathbb{I}$ is the identity operator for the input state space.
Similarly, we define another generalized measurement operator ${\cal M}_2$ for the second MZI at port $x_2$ as
\begin{equation}
	{\cal M}_2 = \sum_{k=0}^4 \frac{1}{2} \left( \left| k \right> + e^{i2\theta_l} \left| k + 2 \right> \right)\ \left< k \right|	,
\end{equation}
where ${\cal M}_{2}'$ is defined so as to satisfy ${\cal M}_2^\dag {\cal M}_2 + {\cal M}_2'^\dag {\cal M}_2' = \mathbb{I}$.
Therefore, the function of the cascaded MZIs at port $x_2$ is given by ${\cal M}_{\textnormal{CMZI}} = {\cal M}_2 {\cal M}_1$.
We postselect a state in which all the input time slots overlap
in order to obtain the probability corresponding to the projection onto Eq. \eqref{eq:Fourier}.
The postselection corresponds to the projection onto $\left| 3 \right>$ at port $x_2$.
Thus, the whole measurement operator ${\cal M}_{\textnormal w}$ for our measurement setup with $d=4$ is given by
\begin{eqnarray}
	{\cal M}_{\textnormal w} &=& \left| 3\right>  \left<3\right| {\cal M}_{\textnormal{CMZI}}	\notag	\\
		&=& \frac{1}{4} \left| 3 \right> \left( e^{i3\theta_l} \left<0\right| 
			+ e^{i2\theta_l} \left<1\right| +e^{i\theta_l} \left<2\right| + \left<3\right| \right)	\notag	\\
		&=& \frac{1}{2} e^{i3\theta_l} \left| 3 \right> \left< \theta_l \right|			.	\label{eq:Mw}
\end{eqnarray}
If we have a density operator of input time-bin state $\rho$,
the probability of finding a photon in the fourth time slot at port $x_2$ is given by
$\mathrm{Tr}({\cal M}_{\textnormal w} \rho {\cal M}_{\textnormal w}^\dag) = \left< \theta_l \right| \rho \left| \theta_l \right> /4$.
Thus, we can obtain the probability of projecting the input state onto Eq. (\ref{eq:Fourier}) by extracting photon detection events in the fourth time slot.

Note that if we cascade $n$ MZIs with $T$, $2T$, ... and $2^{n-1}T$ time delays,
we can implement measurements corresponding to Eq. (\ref{eq:Fourier}) with $d = 2^n$.
The multi-arm interferometer reported in \cite{Thew2004} requires $2^n$ optical passes to perform the same measurements,
which makes it harder to stabilize than the cascaded MZIs.

\section{\label{sec:ExpSet}Experimental setup}
\begin{figure*}[htbp]
	\begin{center}
		\includegraphics[clip,width=15.0cm]{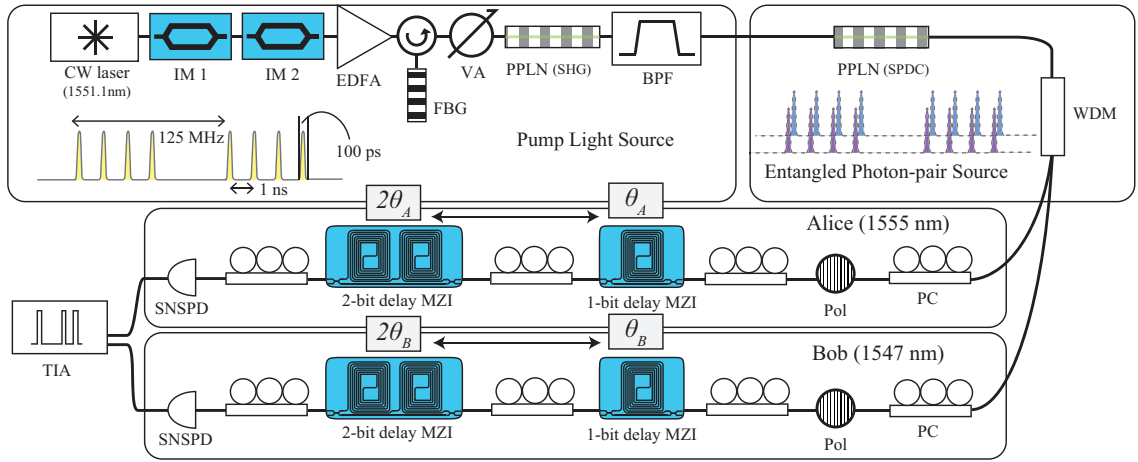}
		\caption{Experimental setup. CW laser: Continuous wave laser. IM 1, IM 2: Lithium niobate intensity modulators.
		EDFA: Erbium-doped fiber amplifier. FBG:  Fiber Bragg grating filter. VA: Variable attenuator. 
		PPLN: Periodically poled lithium niobate waveguide. BPF: Band-pass filter. WDM: Wavelength demultiplexing filter.
		PC: Polarization controller. Pol: Polarizer. 1-bit delay MZI, 2-bit delay MZI: Delay Mach-Zehnder interferometers fabricated using PLC.
		SNSPD: Superconducting nanowire single photon detector. TIA: Time interval analyzer.}
		\label{fig:ExpSetup}
	\end{center}
\end{figure*}
Figure \ref{fig:ExpSetup} shows the experimental setup.
First, we generate four sequential pulses by modulating a CW laser light with a 1551.1-nm wavelength.
The temporal interval, pulse duration, and repetition frequency are 1 ns, 100 ps, and 125 MHz, respectively.
IM 1 is used only when we generate the optimized entangled state,
and IM 2 is used to shape the time slots.
We then amplify the pulse train using an EDFA
and adjust the average power of the pulse train with a VA.
These pulses are launched into a PPLN waveguide, where 780-nm pump pulses are generated via SHG.
The 780-nm pump pulses are input into another PPLN waveguide
so that we can generate time-bin-entangled ququarts whose state is expressed as Eq. (\ref{eq:MaxEnt}) or (\ref{eq:OptEnt}) through SPDC.
The generated photon pairs are separated into channels A and B by wavelengths
using a WDM.
The wavelengths of channels A and B are 1555 and 1547 nm, respectively,
both with a 100-GHz bandwidth.
Photons A and B are sent to Alice and Bob, where they perform the measurement denoted by Eq. (\ref{eq:Fourier}) using the cascaded MZIs,
whose output ports are connected to SNSPDs.
Here, we employ 1- and 2-ns delay MZIs, which we call 1- and 2-bit delay MZIs, respectively.
These MZIs are fabricated using PLC and are thus very stable \cite{Takesue2005,Honjo2004}.
The phase shifts of each MZI are precisely controlled by means of the thermooptic effect caused by electrical heaters attached to the waveguides.
The phase shifts of 1- and 2-bit delay MZIs are synchronized so that we can perform the measurement expressed as Eq. (\ref{eq:Fourier}).
Polarization controllers are placed in front of each MZI to operate them for one polarization.
The detection event signals from the SNSPDs are input into a TIA, where coincidence analysis is performed.
The detection efficiencies of the SNSPDs for Alice and Bob are 19 and 17 \%, respectively,
both with a $<10$-cps dark count rate.

\section{\label{sec:Result}Results}
\subsection{\label{subsec:Res_Max}Maximally entangled state}

\begin{figure}[htbp]
	\begin{center}
		\includegraphics[clip,width=9.0cm]{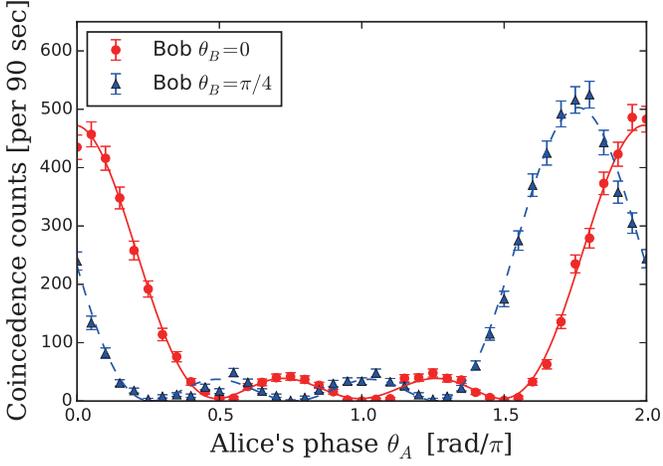}
		\caption{Coincidence counts for the maximally entangled state as a function of phase shifts for Alice's interferometers.}
		\label{fig:MaxFringe}
	\end{center}
\end{figure}
We first generated a maximally entangled state.
Figure \ref{fig:MaxFringe} shows the result when Alice swept her phase of the cascaded MZIs, $\theta _ A$, from $0$ to $2 \pi$
while Bob set his phase, $\theta _ B$, at $0$ or $\pi / 4$.
Circles and triangles are measured coincidence counts for $\theta _ B = 0$ and $\pi / 4$, respectively,
and solid and dotted lines show the fitted curves obtained with the Levenberg-Marquardt algorithm.
The average photon pair number per ququart was set at 0.01,
and measurement time for each phase combination was 90 sec.
Note that throughout this work we did not subtract any noise counts, including accidental coincidence counts due to detector dark counts or multi-photon emissions.
We observed clear coincidence fringes, which are different from the cosine-curve fringes obtained with entangled qubits and unique to entangled ququarts.
In addition, we observed a clear correlation for two non-orthogonal bases at Bob's measurement,
which indicated the existence of a non-classical correlation between ququarts.
For an ideal maximally entangled state, the coincidence count probability is given by
\begin{eqnarray}
	P_{\textnormal{MES}}(\theta _A, \theta _ B) &=& \left| \left< \theta_A \right| \left< \theta_B \right| \left. \Psi_{MES} \right >	\right|^2	\notag	\\
		& = & \frac{1}{4} \cos^2 \left(\frac{\theta _ A + \theta _ B}{2} \right) \cos^2 \left(\theta _ A + \theta _ B \right)	.	\label{eq:CoMaxIdeal}
\end{eqnarray}
Therefore, the measured coincidence count in the experiment can be fitted with
\begin{equation}
	C_{\textnormal{MES}} ^{\textnormal{Fit}}(\theta _A, \theta _ B) = m_1 \frac{ P_{\textnormal{MES}}(\theta _A, \theta _ B) }{\varDelta P_{\textnormal{MES}}} + m_2	,
\end{equation}
where $\varDelta P_{\textnormal{MES}} = 1/4$ is the difference between the maximum and the minimum values of $P_{\textnormal{MES}}$, ans $m_1, m_2$ are fitting parameters.
The visibility of the coincidence fringe is given by 
\begin{equation}
	V = \frac{m_1}{m_1 + 2 m_2}	.		\label{eq:VisMes}
\end{equation}
From the fitted curves, we obtained $V$ of $98.25 \pm 0.86$ and $99.96 \pm 0.94\%$ for $\theta _ B = 0$ and $\pi/4$, respectively.
If we assume symmetric noise, the depolarized mixed state is given by
\begin{equation}
	\rho = \lambda \left | \Psi _ {\textnormal{MES}} \right > \left < \Psi _ {\textnormal{MES}} \right | + \frac{1-\lambda}{16} \mathbb{I}_{16}	,
\end{equation}
where $\lambda$ is a mixing parameter and $\mathbb{I}_{16}$ is the identity operator for the entangled ququarts space.
From this assumption,
the visibility is related to $\lambda$ with the following equation: 
\begin{equation}
	V = \frac{16 \varDelta P_{\textnormal{MES}} \lambda }{2 + \lambda \left( 16 \varDelta P_{\textnormal{MES}} -2 \right)}		.	\label{eq:VisLam}
\end{equation}
If $\lambda > 0.69055$, the state $\rho$ can violate the CGLMP inequality.
Thus a visibility of $81.7\%$, obtained by substituting $\lambda = 0.69055$ into Eq. (\ref{eq:VisLam}) ,
is a critical limit for a coincidence fringe of the maximally entangled state to violate the CGLMP inequality.
Our results---$V=98.25 \pm 0.86$ and $99.96 \pm 0.94 \%$---are obviously larger than $81.7 \%$.

\begin{figure}[t]
	\begin{center}
		\includegraphics[clip,width=9.0cm]{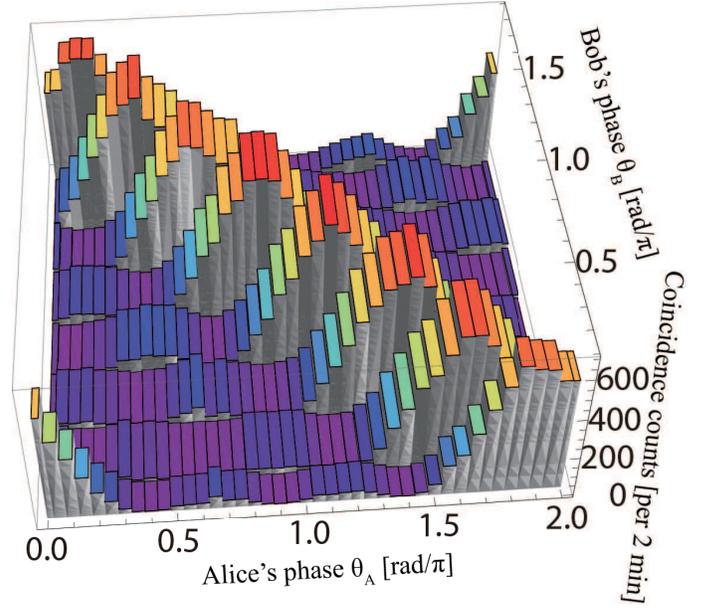}
		\caption{Coincidence counts for the maximally entangled state as a function of phase shifts for Alice and Bob's interferometers.}
		\label{fig:3DFringe}
	\end{center}
\end{figure}
We also performed an $S_4$ measurement for the CGLMP inequality test by performing the coincidence measurements for 64 combinations of $\theta _ A$ and $\theta _ B$.
Figure \ref{fig:3DFringe} shows coincidence measurements
when $\theta _ A$ and $\theta _ B$ were set at 41 and 8 points in their phase range $[0,2\pi]$ and $[\pi/8, 15\pi/8]$, respectively.
The average photon pair number per ququart was set at 0.01,
and the measurement time for each phase combination was 120 sec.
The coincidence counts for the maximally entangled state in the $S_4$ measurement for the CGLMP inequality test
are shown in Table \ref{Tab:RawDataMES} in Appendix \ref{Ap:RawData}.
With those data,
we obtained $S_4$ of $2.774 \pm 0.025$ where the error was estimated by assuming Poisson statistics.
This value violates the CGLMP inequality by 31 standard deviations.

\subsection{\label{subsec:Res_Opt}Optimized entangled state}
\begin{figure}[t]
	\begin{center}
		\includegraphics[clip,width=9.0cm]{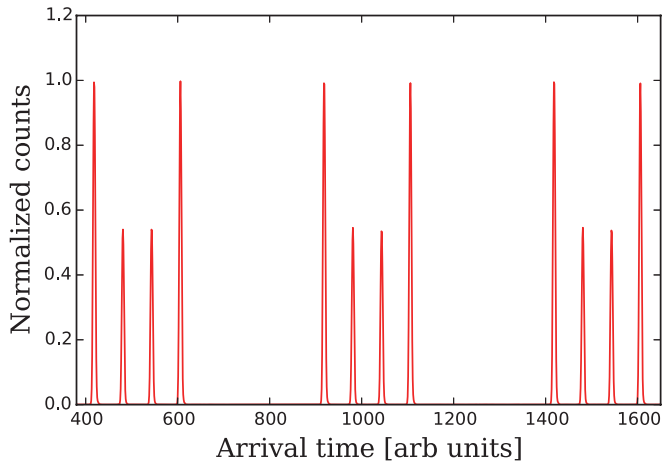}
		\caption{Normalized histogram of single counts for the optimized entangled state at channel A measured after the WDM in Fig. \ref{fig:ExpSetup}.}
		\label{fig:OptSingle}
	\end{center}
\end{figure}
To generate the optimized entangled state,
we modulated the CW light with IM 1 in Fig. \ref{fig:ExpSetup} so that the generated time-bin-entangled ququarts satisfied the condition in Eq. (\ref{eq:OptCondition}).
IM 2 was used to shape the light into four sequential pulses.
Figure \ref{fig:OptSingle} shows a histogram of the photon detection at channel A measured after the WDM filter.
The four peaks correspond to the probabilities of finding a photon in time slots 0, 1, 2, and 3, respectively.
From the histogram, we confirmed the experimental amplitude modulation factor $\gamma_{exp} = 0.738$,
which was very close to the target value (0.739).

\begin{figure}[b]
	\begin{center}
		\includegraphics[clip,width=9.0cm]{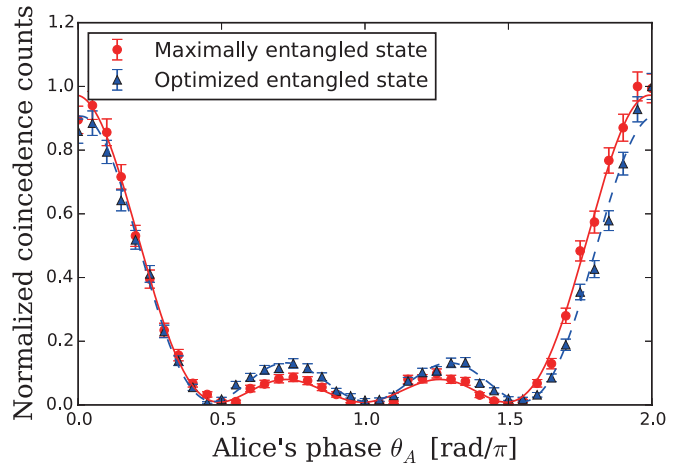}
		\caption{Normalized coincidence counts for the maximally entangled state and the optimized entangled state
		as a function of phase shifts for Alice's interferometers when Bob set his phase at 0.}
		\label{fig:OptFringe}
	\end{center}
\end{figure}
We performed a coincidence fringe measurement.
Triangles in Fig. \ref{fig:OptFringe} show the normalized coincidence as a function of $\theta_A$
when $\theta_B$ was set at 0.
The average photon pair number per ququart was set at 0.02,
and the measurement time for each phase setup combination was 60 sec.
The result for the maximally entangled state when Bob set his phase at 0 is also plotted for comparison (denoted by circles).
The smaller two peaks are enhanced compared with those for the maximally entangled state,
which is a feature of the coincidence fringe shape for the optimized entangled state.
For an ideal optimized entangled state, the coincidence count probability is given by
\begin{widetext}
	\begin{eqnarray}
		P_{\textnormal{OES}}(\theta _A, \theta _ B)  = \left| \left< \theta_A \right| \left< \theta_B \right| \left. \Psi_{\textnormal{OES}} \right >	\right|^2	
			 =  \frac{1}{8(1+\gamma^2)} \left\{ \cos{\frac{3}{2}\left(\theta _ A + \theta _ B \right)} + \gamma \cos \left({\frac{\theta _ A + \theta _ B}{2}} \right) 	\right\}^2	.
	\end{eqnarray}
\end{widetext}
Therefore, the measured coincidence count in the experiment can be fitted by
\begin{equation}
	C_{\textnormal{OES}} ^{\textnormal{Fit}}(\theta _A, \theta _ B) = m_1 \frac{ P_{\textnormal{OES}}(\theta _A, \theta _ B) }{\varDelta P_{\textnormal{OES}}} + m_2	,
\end{equation}
where
\begin{equation}
	\varDelta P_{\textnormal{OES}} = \frac{(1+\gamma)^2}{8(1+\gamma^2)}	,
\end{equation}
and the visibility is given by Eq. (\ref{eq:VisMes}).
From the fitted curves, we obtain $V =97.72 \pm 1.30$ and $97.44 \pm 0.96 \%$ for $\theta_B = 0$ and $\pi/4$, respectively.
Estimating the critical limit with a similar procedure used for the maximally entangled state,
we obtain the limit of the visibility $V = 80.1 \%$.
Thus, the visibilities obtained in our experiment were larger than the critical limit for violating the CGLMP inequality.

Finally, we performed an $S_4$ measurement for the CGLMP inequality test with the optimized entangled state.
The average photon pair number per ququart was set at 0.01,
and the measurement time for each phase setup combination was 120 sec.
As a result, we obtained $S_4$ of $2.913 \pm 0.023$,
which violated the CGLMP inequality by 39 standard deviations
(the coincidence count data are shown in Table \ref{Tab:RawDataOES} in Appendix \ref{Ap:RawData}).
Thus, we successfully confirmed that the use of the optimized entangled state leads to clear enhancements of
both the value of $S_4$ and the violation of the classical limit compared with those obtained with the maximally entangled state.

Unfortunately, there were discrepancies between our results and the theoretical maximum $S_4$ in Table \ref{Tab:Vio_Max_Opt}.
The effects of accidental coincidences caused by the detector dark counts were negligibly small.
The effect caused by multi-photon emission would be a main source of the deviation (see Appendix \ref{Ap:MultiPhoton}).
Another source of the deviation was phase drift due to variations in the CW laser frequency and MZIs settings
during the long measurements required for the coincidence measurements with many phase combinations.
In addition,
the loss difference between the long and short arms of the MZIs may have resulted in the non-ideal measurements for the ququarts.
The accumulation of these experimental defects caused the deviations of the measured values from the theoretical ones.

\section{\label{sec:Conc}Conclusion}
We have described a CGLMP inequality test using optimized time-bin ququarts.
We generated the optimized entangled ququarts by modulating the pump pulse amplitudes
and implemented the $S_4$ measurement for the CGLMP inequality test with cascaded MZIs.
We observed Bell-type parameters $S_4 = 2.774 \pm 0.025$ and $2.913 \pm 0.023$ for the maximally entangled state and the optimized entangled state, respectively.
As a result, we observed the enhancement of both the value of $S_4$ and the violation of the CGLMP inequality with the optimized entangled ququarts.
We hope that our result will lead to advanced quantum communications systems such as high-dimensional robust QKD systems.

\acknowledgements
We thank K. Inoue, K. Igarashi, N. Matsuda, and F. Morikoshi for fruitful discussions.

\appendix

\section{\label{Ap:RawData}Coincidence count data for the $S_4$ measurements}
Here, we show the coincidence count results for the $S_4$ measurements.
Tables \ref{Tab:RawDataMES} and \ref{Tab:RawDataOES} show the raw coincidence count data for the maximally entangled state and the optimized entangled state, respectively.
The details of the measurement conditions are described in Sec. \ref{sec:Result}.
From these results, we calculated the coincidence probabilities and obtained $S_4$ by substituting them into Eq. (\ref{eq:SValueDefinition}).

\begin{table}[htb]
  \begin{center}
  \caption{Coincidence counts for $\left | \Psi _ {\textnormal{MES}} \right>$.}
  \begin{tabular}{c|cc|cccc|cccc|} 	
   \multicolumn{3}{c}{} & \multicolumn{8}{c}{Bob}		\\	\cline{4-11}
    \multicolumn{3}{c|}{} & \multicolumn{4}{c|}{Basis 0} & \multicolumn{4}{c|}{Basis 1}	\\
    \multicolumn{3}{c|}{} & $\frac{\pi}{8}$ & $-\frac{3}{8}\pi$ & $-\frac{7}{8}\pi$ & $-\frac{11}{8}\pi$ & $-\frac{\pi}{8}$ & $-\frac{5}{8}\pi$ & $-\frac{9}{8}\pi$ & $-\frac{13}{8}\pi$		\\	\cline{2-11}
    & & 0			&	605&	72&	34&	49&	493&	36&	37&	67	\\	
    &Basis & $\frac{\pi}{2}$&	46&	453&	74&	17&	62&	545&	31&	38	\\	
    &0 & $\pi$		&	29&	40&	508&	85&	30&	45&	555&	27	\\	
Alice& & $\frac{3}{2}\pi$	&	102&	32&	33&	535&	26&	26&	48&	671	\\	\cline{2-11}
    & & $\frac{\pi}{4}$	&	102&	529&	40&	47&	515&	94&	23&	53	\\	
    &Basis & $\frac{3}{4}\pi$&	30&	28&	473&	28&	22&	445&	92&	24	\\	
    &1 & $\frac{5}{4}\pi$	&	47&	15&	97&	581&	25&	28&	581&	98	\\	
    & & $\frac{7}{4}\pi$	&	611&	22&	18&	48&	67&	27&	34&	600	\\	\cline{2-11}
   \end{tabular}
   \label{Tab:RawDataMES}
   \end{center}
\end{table}

\begin{table}[htb]
  \begin{center}
  \caption{Coincidence counts for $\left | \Psi _ {\textnormal{OES}} \right>$.}
  \begin{tabular}{c|cc|cccc|cccc|} 	
   \multicolumn{3}{c}{} & \multicolumn{8}{c}{Bob}		\\	\cline{4-11}
    \multicolumn{3}{c|}{} & \multicolumn{4}{c|}{Basis 0} & \multicolumn{4}{c|}{Basis 1}	\\
    \multicolumn{3}{c|}{} & $\frac{\pi}{8}$ & $-\frac{3}{8}\pi$ & $-\frac{7}{8}\pi$ & $-\frac{11}{8}\pi$ & $-\frac{\pi}{8}$ & $-\frac{5}{8}\pi$ & $-\frac{9}{8}\pi$ & $-\frac{13}{8}\pi$		\\	\cline{2-11}
    & & 0			&	544&	38&	21&	60&	517&	54&	33&	30	\\	
    &Basis & $\frac{\pi}{2}$&	57&	426&	46&	24&	24&	458&	47&	53	\\	
    &0 & $\pi$		&	29&	63&	470&	25&	20&	26&	453&	53	\\	
Alice& & $\frac{3}{2}\pi$	&	30&	49&	63&	408&	57&	43&	21&	445	\\	\cline{2-11}
    & & $\frac{\pi}{4}$	&	57&	462&	64&	42&	517&	40&	18&	84	\\	
    &Basis & $\frac{3}{4}\pi$&	52&	29&	422&	35&	57&	398&	44&	20	\\	
    &1 & $\frac{5}{4}\pi$	&	70&	28&	51&	439&	56&	80&	430&	31	\\	
    & & $\frac{7}{4}\pi$	&	459&	55&	48&	30&	44&	40&	71&	408	\\	\cline{2-11}
   \end{tabular}
   \label{Tab:RawDataOES}
   \end{center}
\end{table}

\section{\label{Ap:MultiPhoton}Effect caused by multi-photon emission}
Here, we estimate the degradation of the visibility and $S_d$ caused by the multi-photon emission.
In our experiment, the coherence time of a single photon (10 ps) was much smaller than the pump pulse duration.
Thus, the multiple photons can be approximated to distinguishable photons.
It is known that the number of such a distinguishable photon pair generated via SPDC follows a Poisson distribution \cite{Takesue2010a}.
Therefore, we can estimate the coincidence probability $P_{co}$ as follows.
From the Poisson statistics, the probability of generating $m$ photon pairs $P_{po}$ is given by 
\begin{equation}
	P_{po} = \frac{\mu^m}{m!} \exp\left( -\mu \right)	,	\label{eq:Poisson}
\end{equation}
where $\mu$ is a average number of photon pairs.
When we generate $N$ photon pairs ($m=N$),
the probability of losing $(N-n)$ photons at Alice's side because of the medium loss, $P_{loss A}$, is given by
\begin{equation}
	P_{loss A} = {}_{N} \mathrm{C} _{n} \eta^{n} \left( 1-\eta \right)^{N-n}		,	\label{eq:LossBinomial}
\end{equation}
where $\eta$ is a transmittance including detector efficiency.
When Alice receives $n$ photons, the probability of Alice detecting $n_A$ photons in her measurement setup, $P_{det A}$, is given by
\begin{equation}
	P_{det A} = {}_{n} \mathrm{C} _{n_A} p_A^{n_A} \left( 1-p_A \right)^{n-n_A}	,	\label{eq:AliceDetBinomial}
\end{equation}
where
\begin{equation}
	p_A = \mathrm{Tr} \left( {\cal M}_A \rho {\cal M}_A^\dag \right)	.
\end{equation}
$\rho$ and ${\cal M}_A$ are a density operator of a single photon pair and a generalized measurement operator for Alice's measurement setup, respectively.
For example, $\rho = \left | \Psi _ {\textnormal{OES}} \right > \left < \Psi _ {\textnormal{OES}} \right |$
and ${\cal M}_A = {\cal M}_{w}$ for our coincidence measurement with the optimized entangled state.

Here,
we define probabilities $p_{B|A}$, $p_{B|\overline{A}}$, and $p_{B}$ as follows.
\begin{eqnarray}
	p_{B|A} &=& \frac{p_{A,B}}{p_A}	,	\\
	p_{B|\overline{A}} &=& \frac{p_B - p_{A,B}}{1-p_A}	,	\\
	p_{B} &=& \mathrm{Tr} \left( {\cal M}_B \rho {\cal M}_B^\dag \right)	,	\\
	p_{A,B} &=& \mathrm{Tr} \left( {\cal M}_B {\cal M}_A \rho {\cal M}_A^\dag {\cal M}_B^\dag \right)	,
\end{eqnarray}
where ${\cal M}_B$ is a generalized measurement operator for Bob.
The pairs of photons detected by Alice are detected by Bob with a probability $\eta p_{B|A}$.
The pairs of photons that are not lost in transmission and not detected in Alice's measurement setup are detected by Bob with a probability $\eta p_{B|\overline{A}}$.
The pairs of photons lost in transmission are detected by Bob with a probability $\eta p_{B}$.
Our SNSPD cannot resolve the number of the detected photons,
so the coincidence probability on the above conditions, $p_{co}'$, is given by
\begin{equation}
	p_{co}' = 1 - \left( 1- \eta p_{B|A} \right)^{n_A}	\left( 1-\eta p_{B|\overline{A}} \right)^{n-n_A}	\left( 1-\eta p_{B} \right)^{N-n}	.	\label{eq:ConditionalCoinc}
\end{equation}
By utilizing Eqs. (\ref{eq:Poisson}) -- (\ref{eq:AliceDetBinomial}) and (\ref{eq:ConditionalCoinc}),
the coincidence probability is given by
\begin{equation}
	P_{co} = \sum_{N=1}^{\infty} P_{po} \sum_{n=1}^{N} P_{loss A} \sum_{n_A=1}^{n} P_{det A} p_{co}'	.	\label{eq:PcoBeforeCal}
\end{equation}

By calculating the summation in Eq. (\ref{eq:PcoBeforeCal}), we obtain
	\begin{equation}
		P_{co} 
			= 1 - e^{-\mu \eta p_A } - e^{-\mu \eta p_B } + e^{-\mu \eta \left(p_A + p_B - \eta p_{A,B} \right) }		.
				\label{eq:PcoAfterCal}
	\end{equation}
When $\eta \ll 1$, Eq. (\ref{eq:PcoAfterCal}) is approximated as
\begin{equation}
	P_{co} \approx \mu \eta^2 \left(p_{A,B} + \mu p_{A}p_{B} \right)	.	\label{eq:PcoAprrox}
\end{equation}
For example, when we use the optimized entangled ququarts and cascaded MZIs,
probabilities $p_A$, $p_B$, and $p_{A,B}$ are given by
\begin{eqnarray}
	p_A &=& p_B = \frac{1}{16}	,	\\
	p_{A,B} &=& \frac{1}{16} P_{\textnormal{OES}}(\theta _A, \theta _ B)	.
\end{eqnarray}
By substituting these probabilities into Eq. (\ref{eq:PcoAprrox}) and using the definition of visibility, we obtain the visibility as
\begin{equation}
	V = \frac{(1+\gamma)^2}{(1+\gamma)^2+\mu(1+\gamma^2)}	.	\label{eq:Multi-photonVis}
\end{equation}
We combine Eq. (\ref{eq:Multi-photonVis}) and the visibility for the optimized entangled ququarts as a function of the mixing parameter $\lambda$ as Eq. (\ref{eq:VisLam}),
and $\lambda$ is given by
\begin{equation}
	\lambda = \frac{1}{1+\mu}	.
\end{equation}
We obtain $\lambda = 0.990$ for $\mu=0.01$.
This predicts $S_4$ of $2.943$,
which is close to our experimental result of $S_4 = 2.913$.


\bibliographystyle{prX_Modoki.bst}
\bibliography{library}

\end{document}